\documentclass[12pt,a4paper]{article}

\usepackage{bm}

\usepackage{amsmath}
\usepackage{amsfonts}
\usepackage{amssymb}

\newcommand{\CC}{\mathbb{C}}
\newcommand{\ZZ}{\mathbb{Z}}
\newcommand{\RR}{\mathbb{R}}

\newcommand{\ol}{\overline}
\newcommand{\wt}{\widetilde}

\newcommand{\BPS}{\text{-BPS}}

\def\tr{\mathop{\rm tr}\nolimits}
\def\Pf{\mathop{\rm Pf}\nolimits}
\def\diag{\mathop{\rm diag}\nolimits}
\def\Pexp{\mathop{\rm Pexp}\nolimits}

\newcommand{\sX}{\mathsf{X}}
\newcommand{\sY}{\mathsf{Y}}
\newcommand{\sZ}{\mathsf{Z}}

\begin{document}

%titlepage
\begin{titlepage}
\title{
\vspace{-1.5cm}
\begin{flushright}
{\normalsize TIT/HEP-670\\ January 2019}
\end{flushright}
\vspace{3.5cm}
\Huge{BPS Partition Functions for S-folds}}
\author{\large{
Reona {\scshape Arai\footnote{E-mail: r.arai@th.phys.titech.ac.jp}},
Shota {\scshape Fujiwara\footnote{E-mail: s.fujiwara@th.phys.titech.ac.jp}},
and 
Yosuke {\scshape Imamura\footnote{E-mail: imamura@phys.titech.ac.jp}}
}\\
\\
{\itshape Department of Physics, Tokyo Institute of Technology}, \\ {\itshape Tokyo 152-8551, Japan}}

\date{}
\maketitle
\thispagestyle{empty}

%abstract
\begin{abstract}
We derive a formula for the BPS partition functions
of arbitrary S-fold theories.
We first generalize the known result for the ${\cal N}=4$ $U(N)$ supersymmetric
Yang-Mills theory
to $SO$ and $Sp$ theories,
and then we extend the formula
to ${\cal N}=3$ theories.
We confirm that the results for rank $1$ and $2$ are consistent to
the supersymmetry enhancement from ${\cal N}=3$ to ${\cal N}=4$.
We also derive the same formula from the quantization of
D3-branes in $\bm{S}^5/\ZZ_k$.
\end{abstract}
\end{titlepage}
\section{Introduction}
Four-dimensional superconformal field theories
(SCFT) have been studied for many years.
We have learned a lot especially in $\mathcal{N}=1,2$, and $4$ cases.
However, $\mathcal{N}=3$ theories are not well understood.
This is because genuine $\mathcal{N}=3$ theories, 
which do not have hidden $\mathcal{N}=4$ supersymmetry, 
have no Lagrangian description and realized only at the strong coupling regime.

In recent years, some progress has been made toward understanding of the theories. 
Aharony and Evtikhiev \cite{Aharony:2015oyb}
derived some universal properties
of $\mathcal{N}=3$ theories
from arguments based on
the $\mathcal{N}=3$ superconformal algebra.
In particular, they showed that genuine $\mathcal{N}=3$ theories
cannot have marginal deformations.
See also \cite{Cordova:2016xhm} for more comprehensive
analysis of marginal deformations.
The absence of marginal deformations is
consistent to the well-known fact that
the only ${\cal N}=3$ free field multiplet
is the vector multiplet, which is after the CPT completion
equivalent to the ${\cal N}=4$ vector multiplet.
Therefore, a construction of a genuine ${\cal N}=3$ theory
is necessarily non-perturbative.

A class of ${\cal N}=3$ theories were constructed in
\cite{Garcia-Etxebarria:2015wns} as the theories on
D3-branes in S-fold backgrounds.
(See also \cite{Ferrara:1998zt} for a construction of related supergravity backgrounds.)
S-folds are generalization of the orientifold.
From the viewpoint of F-theory,
the orientifold with $3+1$ dimensional fixed plane
(O3-plane) is a $\ZZ_2$ orbifold accompanied by the rotation
of the toric fiber by angle $\pi$.
If the modulus $\tau$ of the toric fiber takes one
of the special values $\tau=\exp(2\pi i/k)$ ($k=3,4,6$),
we can define the S-fold as the $\ZZ_k$ orbifold
of the spacetime accompanied by the $2\pi/k$ rotation of the
toric fiber.
This leaves $24$ supersymmetries unbroken, and
if we put D3-branes parallel to the
fixed plane, a four-dimensional ${\cal N}=3$ theory
is realized on the worldvolume.
We call them S-fold theories.

An S-fold theory is specified by the order $k$ of
the S-fold group $\ZZ_k$, the rank $N$, which is the number of D3-branes,
and, in addition, an integer $p$ associated with the discrete torsion of
the NS-NS and the R-R three-form fluxes \cite{Garcia-Etxebarria:2015wns,Witten:1998xy,Aharony:2016kai,Imamura:2016abe}.
We denote the theory by $S(k,N,p)$.

Because S-fold theories are defined by using the brane construction,
it is natural to analyze them with
the AdS/CFT correspondence \cite{Maldacena:1997re}.
For the ${\cal N}=4$ $SU(N)$ supersymmetric Yang-Mills theory (SYM)
it was first pointed out in \cite{Witten:1998qj} that
there is a one-to-one correspondence
between BPS operators in the SYM and Kaluza-Klein (KK) modes
in $AdS_5\times \bm{S}^5$ \cite{Kim:1985ez,Gunaydin:1984fk},
and the agreement of the superconformal indices and the BPS partition functions
calculated on the two sides of the duality
were demonstrated in \cite{Kinney:2005ej}.
For S-folds, the gravity background is replaced by
$AdS_5\times \bm{S}^5/\ZZ_k$,
and the superconformal index in the large $N$ limit
was obtained by a simple $\ZZ_k$ projection in \cite{Imamura:2016abe}.
A relation between wrapped branes around non-trivial cycles in $\bm{S}^5/\ZZ_k$
and gauge invariant operators was also discussed in \cite{Aharony:2016kai,Imamura:2016abe}.

In this paper we focus on BPS partition functions of S-fold theories.
Particularly, we construct grand partition functions for ${\cal N}=3$ theories
associated with operators made of
the scalar fields $(\mathsf{X},\mathsf{Y},\mathsf{Z})$.
The Taylor expansion of these grand partition functions generates
the partition functions for an arbitrary $N$.

The gauge invariant operators made of the scalar fields
form the coordinate ring of the moduli space.
The moduli spaces of the S-fold theories are
all orbifolds of the form ${\cal M}_{S(k,N,p)}=\CC^{3N}/{\cal W}_{S(k,N,p)}$.
There is a general formula known as ``the Molien series''
which gives the partition function for an arbitrary orbifold.
See \cite{Benvenuti:2006qr} for an application to the theories on D3-branes probing orbifolds.
The Molien series is also applicable to S-fold theories,
and in this sense our result is not novel.
However, the formula we derive in this paper
has very simple structure which cannot be seen in the
Molien series, and the correspondence to the
holographic picture is clearer.
In particular, the formula gives separately the contribution of
each sector specified by the D3-brane winding number in the internal space
$\bm{S}^5/\ZZ_k$.
We show that the same formula is reproduced from the analysis of D3-branes.

This paper is organized as follows.
In the next section we review the BPS partition function of
the $\mathcal{N}=4$ $U(N)$ SYM following \cite{Kinney:2005ej}.
The resulting partition function is the same as that
of $N$ bosonic particles in a three-dimensional harmonic potential.
In section \ref{sosp.sec} we construct grand partition functions
for ${\cal N}=4$ SYM theories with $SO$ and $Sp$ gauge groups realized by orientifolds.
These can be regarded as S-folds with $k=2$.
In section \ref{sfolds.sec} we generalize the results
of section \ref{sosp.sec} and construct grand partition functions
for $k=3,4$, and $6$ theories.
We confirm that the partition function is consistent to
the supersymmetry enhancement proposed in \cite{Nishinaka:2016hbw,Aharony:2016kai}.
We also comment on some relations among partition functions via discrete gaugings.
In section \ref{d3.sec} we reproduce the same formula by quantizing D3-branes
in $\bm{S}^5/\ZZ_k$.
Finally, in section \ref{discussion.sec} we discuss our results and open questions.
In appendix \ref{MS} we give the definition of the Molien series and its application to the $U(N)$ theory.

%%%%%%%%%%%%%%%%%%%%%%%%%%%%%%%%%%%%%%%%%%%%%%%%%%%
\section{BPS partition functions of $U(N)$ SYM}

In this section we review
the BPS partition function for the ${\cal N}=4$
$U(N)$ SYM ,
which has been well understood,
and define some notations which will be used in the
following sections.
Analysis in this section is based mainly on \cite{Kinney:2005ej}.
See also \cite{Benvenuti:2006qr} for a generalization to
a large class of theories associated with Calabi-Yaus.

\subsection{BPS partition function}
The BPS partition function of an ${\cal N}=4$
SYM 
is defined by
\begin{align}
Z(x,y,z)=\tr (x^{J_1}y^{J_2}z^{J_3}),
\label{pf}
\end{align}
where the trace is taken over gauge invariant BPS operators
consisting of the adjoint scalar fields $\mathsf{X}$, $\mathsf{Y}$, and $\mathsf{Z}$.
$J_1$, $J_2$, and $J_3$ are Cartan generators of $SU(4)_R$.
We use an $so(6)$ basis for these generators.
Namely, 
$J_1$, $J_2$, and $J_3$
count the numbers of
$\mathsf{X}$, $\mathsf{Y}$, and $\mathsf{Z}$, respectively.

There are several types of BPS operators.
Operators consisting of one of the scalar fields,
say, $\mathsf{Z}$, preserve half of $16$ supersymmetries,
and are called $\frac{1}{2}$-BPS operators.
In the context of the ${\cal N}=2$ subalgebra
$\mathsf{Z}$ is regarded as the scalar component of
the ${\cal N}=2$ vector multiplet,
and the $\frac{1}{2}$-BPS operators are called
Coulomb branch operators.
The corresponding partition function is often called the Coulomb branch Hilbert series.

BPS operators consisting of $\mathsf{X}$ and $\mathsf{Y}$
preserve quarter supersymmetry, and
are called $\frac{1}{4}$-BPS operators.
$\frac{1}{4}$-BPS operators are also called Higgs branch operators
because $\mathsf{X}$ and $\mathsf{Y}$ belong
to the ${\cal N}=2$ hypermultiplet,
and operators made of them parameterize the Higgs branch.
The corresponding partition function is often called the Higgs branch Hilbert series.

The most general BPS operators consisting of
the three scalar fields
$\mathsf{X}$, $\mathsf{Y}$, and $\mathsf{Z}$
preserve
only two supersymmetries, and are called
$\frac{1}{8}$-BPS operators.

Although all gauge invariant operators made of
$\mathsf{X}$, $\mathsf{Y}$, and $\mathsf{Z}$
are BPS when $g_{\rm YM}=0$,
some of them become non-BPS
if we turn on the coupling constant.
In this paper we are interested in BPS operators
in theories with non-vanishing coupling constant.

If we take the trace over $\frac{1}{2}$ ($\frac{1}{4}$, $\frac{1}{8}$)
BPS operators in (\ref{pf}),
the partition function is called the
$\frac{1}{2}$ ($\frac{1}{4}$, $\frac{1}{8}$, respectively)
BPS partition function.
Because $\frac{1}{2}$ and $\frac{1}{4}$ BPS operators form
subsets of $\frac{1}{8}$-BPS operators,
once we obtain the $\frac{1}{8}$-BPS partition function,
the $\frac{1}{2}$ and $\frac{1}{4}$ BPS partition functions are
obtained by the following specializations:
\begin{align}
Z^{\frac{1}{2}\text{-BPS}}(z)&=
Z(x,y,z)|_{x=y=0}, &
Z^{\frac{1}{4}\text{-BPS}}(x,y)&=
Z(x,y,z)|_{z=0}.
\label{specializations}
\end{align}

In the following,
we describe BPS operators in two ways.
The first is ``the Casimir representation''
in which we represent operators as
polynomials of
trace operators (and Pfaffian operators in some cases).
The other is ``the oscillator representation''
in which we give gauge invariant operators
as polynomials of diagonal components of
scalar fields.
The former is convenient to describe $\frac{1}{2}$-BPS operators
and we can easily calculate the $\frac{1}{2}$-BPS partition function
with this representation
for an arbitrary gauge group.
The latter is suitable to describe more general BPS operators
and enable us to calculate the $\frac{1}{8}$-BPS partition function
of the $U(N)$ SYM.

%%%%%%%%%%%%%%%%%%%%%%%%%%%%%%%%%%%%%%%%%%%%%%%%%%%%%
\subsection{$\frac{1}{2}$-BPS partition function}
We first discuss the $\frac{1}{2}$-BPS sector.
An operator in the $\frac{1}{2}$-BPS sector consists only of one adjoint scalar field $\mathsf{Z}$.

In general
the Coulomb branch chiral ring of an ${\cal N}=4$ SYM
is freely generated
by the Casimir operators made of the scalar field $\mathsf{Z}$.
The number of independent Casimir operators is the same as the
rank of the gauge group.
Let ${\cal O}_i$ ($i=1,\ldots,r$) be the independent
Casimir operators and $d_i$ be their scaling dimensions.
The $\frac{1}{2}$-BPS partition function is given in terms of $d_i$ by
\begin{align}
Z^{\frac{1}{2}\text{-BPS}}(z)
=\prod_{i=1}^r\frac{1}{1-z^{d_i}}
=\Pexp\left(\sum_{i=1}^r z^{d_i}\right),
\label{halfbpsz}
\end{align}
where $\Pexp$ is the plethystic exponential defined by
\begin{align}
\Pexp \left ( f(x_i) \right)=\exp\left(\sum_{m=1}^\infty\frac{1}{m}f(x_i^m)\right).
\end{align}
This is a general formula applicable to the ${\cal N}=4$
SYM with an arbitrary gauge group.
The dimensions $d_i$ for some groups are shown in
Table \ref{Casimir1}.

\begin{table}[htb]
\caption{The dimensions $d_i$ of generating Casimir operators for gauge groups $G$}\label{Casimir1}
\centering
\begin{tabular}{|c|c|}\hline
 $G$  & $d_i$ \\ \hline \hline
$U(N)$ &$1,2,3,\dots, N$  \\ \hline
$SU(N)$ &$2,3,4,\dots, N$  \\ \hline
$SO(2N)$ &$2,4,6,\dots, 2N-2;N$  \\ \hline
$SO(2N+1),Sp(N)$ &$2,4,6,\dots, 2N$  \\ \hline
$G_2$ &$2,6$  \\ \hline
\end{tabular}
\end{table}

Let us consider the $U(N)$ case more explicitly.
The Casimir operators are
\begin{align}
{\cal O}_i=
\tr (\mathsf{Z}^i),\quad
(i=1,2,\ldots ,N).
\label{tracexk}
\end{align}
Operators $\tr (\mathsf{Z}^i)$ with $i\geq N+1$ are not independent of them and
decomposable into smaller traces.
The partition function is
\begin{align}
Z_{U(N)}^{\frac{1}{2}\text{-BPS}}(z)=\prod_{k=1}^N\frac{1}{1-z^k}=\Pexp \left (z+z^2+\cdots +z^N \right ).
\label{zun}
\end{align}

To study the $\frac{1}{4}$ and $\frac{1}{8}$-BPS sectors
the oscillator representation is
more suitable.
Let us first use this representation to the
$\frac{1}{2}$-BPS sector and later generalize it to the $\frac{1}{8}$-BPS
sector.
We diagonalize
the scalar field $\mathsf{Z}$ by the gauge transformation
and let $z_i$ ($i=1,\ldots,N$) be the diagonal components.
$\frac{1}{2}$-BPS gauge invariant operators
are polynomials of these $N$ variables
that are
invariant under the Weyl group ${\cal W}_{U(N)}=S_N$.
We can use the set of following symmetric polynomials as a basis.
\begin{align}
\sum_{\sigma \in S_N}\prod_{i=1}^N z_{\sigma (i)}^{m_i}.
\label{sympoly}
\end{align}
Gauge invariant operators can be regarded as functions
in the moduli space,
and the symmetric polynomials
are generators of the coordinate ring of the
Coulomb branch moduli space
${\cal M}_C=\CC^N/{\cal W}_{U(N)}$.

These basis functions are labeled by a vector of
$N$ non-negative integers
$\{m_1,\ldots ,m_N\}$.
This is the same as the Hilbert space of $N$ one-dimensional
harmonic oscillators.
Two vectors with different orders of the components are identified,
and this is interpreted as the Bose statistics of the particles.
Namely, the $\frac{1}{2}$-BPS partition function of the $U(N)$ theory is identical to
the partition function of $N$ bosonic particles
in the one-dimensional
harmonic potential.

The equivalence of two descriptions, the Casimir representation and
the oscillator representation, becomes obvious if we use
partitions to specify operators.
In the Casimir representation, we can adopt the set of operators
\begin{align}
{\cal O}_{i_1}
{\cal O}_{i_2}
{\cal O}_{i_3}
\cdots
{\cal O}_{i_p},\quad
(N\geq i_1\geq i_2\geq \cdots\geq i_p\geq 0)
\end{align}
labeled by a non-ascending series of integers bounded by $N$ with an arbitrary length
as a basis of gauge invariant operators.
In the oscillator representation,
the symmetric polynomials
(\ref{sympoly}) are labeled by $N$ non-negative integers in non-ascending order:
\begin{align}
m_1\geq m_2\geq \cdots\geq m_N\geq 0.
\end{align}
Elements of both bases are labeled by a non-ascending series of non-negative
integers, each of which can be represented as a Young diagram.
The two Young diagrams
are transposition of each other,
and the two descriptions give the same
partition function.

In the oscillator representation,
it is natural to define the grand partition function
\begin{align}
\Xi_{U(*)}^{\frac{1}{2}\BPS}(z;t) =\sum _{N=0}^{\infty }Z_{U(N)}^{\frac{1}{2}\BPS}(z) t^N.
\end{align}
We use ``$*$'' as a wildcard character to represent the summation
over the rank $N$.
The grand partition function is given as the product of the partition function of
each state of the harmonic oscillator,
\begin{align}
\Xi_{U(*)}^{\frac{1}{2}\BPS}(z;t)
=\prod_{k=0}^{\infty }\frac{1}{1-tz^k}
=\Pexp \left ( \frac{t}{1-z}\right ).
\label{xiun0}
\end{align}
We can easily show
by the $q$-binomial formula
that
(\ref{zun}) is obtained from 
(\ref{xiun0}) by the Taylor expansion with respect to $t$.

In (\ref{xiun0}) $k=0,1,2,\ldots$ labels energy eigenstates
of the harmonic oscillator.
It is important that $k$ starts from zero corresponding to the ground state.
When the energy of the whole system of $N$ particles
is much smaller than $N$,
only small part of $N$ particles are excited,
and the majority of the particles are in the ground state.
The finiteness of $N$ affects
the degeneracy of states only when
the energy is comparable to or greater than $N$.

%%%%%%%%%%%%%%%%%%%%%%%%%%%%%%%%%%%%%%%%%%%%%%%%%%%%%%%%%
\subsection{$\frac{1}{8}$-BPS partition function}
Let us move on to the $\frac{1}{8}$-BPS operators.
In this case we can use
single trace operators
made of $\mathsf{X}$, $\mathsf{Y}$, and $\mathsf{Z}$
as generators of the chiral ring.
However, it is known that unlike the $\frac{1}{2}$-BPS sector
the chiral ring is not freely generated
and there are non-trivial relations
called syzygies \cite{Benvenuti:2006qr}
among the generators, and it is not so easy to
count independent operators in the Casimir representation.
The oscillator representation is more suitable.
Indeed, it is quite easy to give a set of independent operators
in the oscillator representation, and we can find
that they are equivalent to the states
of $N$ bosonic particles in a three-dimensional
harmonic potential \cite{Kinney:2005ej}
as is shortly explained.

Thanks to the $F$-term conditions
$[\mathsf{X},\mathsf{Y}]=[\mathsf{Y},\mathsf{Z}]=[\mathsf{Z},\mathsf{X}]=0$
we can diagonalize the three fields $\mathsf{X}$, $\mathsf{Y}$, and $\mathsf{Z}$
simultaneously by the gauge transformation.
Let $(x_i,y_i,z_i)$ ($i=1,\ldots,N$) be the
diagonal components.
The symmetric polynomial
(\ref{sympoly}) is replaced by
\begin{align}
\sum_ {\sigma \in S_N}\prod _{i=1}^N
\Psi_ {\vec m_i}(x_{\sigma (i)},y_{\sigma (i)},z_{\sigma (i)}),
\label{wavefn}
\end{align}
where
$\Psi _{\vec m}(x,y,z)=x^{m_x}y^{m_y}z^{m_z}$
is the monomial which is labeled by
a three-dimensional vector $\vec m=(m_x,m_y,m_z)$
with non-negative integer components.
As in the $\frac{1}{2}$-BPS case these can be regarded as basis functions of
the coordinate ring of the total moduli space
${\cal M}=\CC^{3N}/{\cal W}_{U(N)}$.

We regard $\Psi _{\vec m}$ as the wave function
of a three-dimensional harmonic oscillator in the state
specified by $\vec m$.
The symmetrization is again interpreted as
the Bose statistics,
and (\ref{wavefn}) can be regarded as the wave function
of $N$ bosonic particles in the three-dimensional harmonic
potential.
The grand partition function is
\begin{align}
\Xi_{U(*)}(x,y,z;t)
=\Pexp \left(tI(x,y,z)\right)
=\prod _{p,q,r=0}^{\infty }\frac{1}{1-x^py^qz^rt},
\label{xiun}
\end{align}
where $I(x,y,z)$ is the function
\begin{align}
I(x,y,z)=\frac{1}{(1-x)(1-y)(1-z)}.
\label{iustar}
\end{align}
By picking up the coefficient of the $t^N$ term from the Taylor expansion of
(\ref{xiun}) we obtain $Z_{U(N)}$, the $\frac{1}{8}$-BPS partition function of
the $U(N)$ theory.
In appendix \ref{MS} we show that the BPS partition functions obtained from (\ref{xiun})
are equal to those from the Molien series.

Interestingly, this partition function can be reproduced
on the gravity side as the contribution of sphere giants \cite{Biswas:2006tj} 
or AdS giants \cite{Mandal:2006tk}.
Descriptions of two types of giant gravitons
are complementary, and each of them gives the result
identical to (\ref{xiun}).

We call the function $I$ in
(\ref{iustar}) ``the single-particle partition function''
by two reasons:
\begin{enumerate}
\item $I$ is the partition function of
a single three-dimensional harmonic oscillator.

\item From the viewpoint of the gravity dual in the large $N$ limit
$I$ can be regarded as the partition function of a
single KK particle in $\bm{S}^5$
(up to the difference by $1$).
\end{enumerate}

Concerning the second reason,
the $N\to \infty$ limit of $Z_{U(N)}$ can be read off from the
pole of $\Xi_{U(*)}$ at $t=1$ as
\begin{align}
Z_{U(\infty )}=\lim_{t\to 1}
\left [(1-t)\Xi _{U(*)} \right ]
=\Pexp \left (I-1\right ).
\label{zuinfty}
\end{align}
``$-1$'' in the last expression eliminates the contribution of the
harmonic oscillator ground state.
Therefore, only the excited states of the harmonic oscillator
correspond to the KK gravitons.

We can define the ``single-particle'' partition function
$I_{U(N)}$ for finite $N$ by $Z_{U(N)}=\Pexp \left( I_{U(N)} \right)$.
For the $\frac{1}{2}$-BPS states, the finite $N$ correction is
given as the simple cut-off at ${\cal O}(z^{N+1})$.
(See (\ref{zun}).)
For $\frac{1}{4}$ and $\frac{1}{8}$-BPS states,
the finite $N$ correction becomes more complicated.

We point out that
the $U(1)$ partition function agrees with the function $I$:
\begin{align}
Z_{U(1)}(x,y,z)=I(x,y,z).
\label{zuone}
\end{align}
We also note that the $SU(N)$ partition function can be obtained by removing
the $U(1)$ factor from the $U(N)$ partition function:
\begin{align}
Z_{SU(N)}(x,y,z)=\frac{Z_{U(N)}(x,y,z)}{Z_{U(1)}(x,y,z)}.
\label{supf}
\end{align}

%%%%%%%%%%%%%%%%%%%%%%%%%%%%%%%%%%%%%%%%%%%%%%%%%%%%%%%%%%%%%%
\section{BPS partition functions for $SO$ and $Sp$ theories}\label{sosp.sec}
\subsection{$\frac{1}{2}$-BPS partition function}\label{halfso}
Let us extend the derivations of BPS partition functions
in the last section to the $SO$ and $Sp$ gauge theories.

We first consider the $\frac{1}{2}$-BPS sector.
For $SO$ theories, the adjoint fields become anti-symmetric matrices,
and the trace operators with odd order become identically zero,
and only ones with even order exist.
In addition, for $SO(2N)$, the Pfaffian operator $\Pf \mathsf{Z}$ joins the
generators of the chiral ring.

Let us first consider the $SO(2N+1)$ theory.
The generators are
\begin{align}
\tr (\mathsf{Z}^2),\quad
\tr (\mathsf{Z}^4),\quad
\cdots,\quad
\tr (\mathsf{Z}^{2N-2}),\quad
\tr (\mathsf{Z}^{2N}).
\label{bngen}
\end{align}
These freely generate the Coulomb branch chiral ring,
and the corresponding partition function is
\begin{align}
Z^{\frac{1}{2}\BPS}_{SO(2N+1)}(z)=\Pexp \left(z^2+\cdots +z^{2N}\right).
\label{zso2np1}
\end{align}
This is obtained simply by replacing $z$ in $Z_{U(N)}$ by $z^2$.
Therefore, the grand partition function is also obtained
from (\ref{xiun0}) by the same replacement:
\begin{align}
\Xi^{\frac{1}{2}\text{-BPS}} _{SO({\rm odd})}(z;t)
=\sum_ {N=0}^{\infty }Z^{\frac{1}{2}\text{-BPS}}_{SO(2N+1)}t^N
=\Pexp \left (\frac{t}{1-z^2}\right ).
\label{soodd2}
\end{align}

In the $SO(2N)$ case
the generators are
\begin{align}
\tr \mathsf{Z}^2,\quad
\tr \mathsf{Z}^4,\quad
\cdots,\quad
\tr \mathsf{Z}^{2N-2},\quad
\Pf \mathsf{Z}.
\label{so2ngen}
\end{align}
Although we can of course directly calculate
$Z^{\frac{1}{2}\BPS}_{SO(2N)}$
by the general formula (\ref{halfbpsz}),
it is instructive to give it
by using (\ref{zso2np1}).
The difference of the list of
generators (\ref{so2ngen}) from (\ref{bngen})
is that
the Pfaffian $\Pf \mathsf{Z}$ joins the generators,
and instead $\tr \mathsf{Z}^{2N}$ becomes a dependent operator which is
decomposable into $\det \mathsf{Z}=(\Pf \mathsf{Z})^2$ and smaller traces.
Correspondingly,
$Z^{\frac{1}{2}\BPS}_{SO(2N)}$
is obtained from $Z^{\frac{1}{2}\BPS}_{SO(2N+1)}$
by
removing the factor $1/(1-z^{2N})$ and introducing the new factor $1/(1-z^N)$:
\begin{align}
Z^{\frac{1}{2}\text{-BPS}}_{SO(2N)}
=\frac{1-z^{2N}}{1-z^N}Z^{\frac{1}{2}\text{-BPS}}_{SO(2N+1)}
=(1+z^N)Z^{\frac{1}{2}\text{-BPS}}_{SO(2N+1)}.
\end{align}
The corresponding grand partition function is
\begin{align}
\Xi ^{\frac{1}{2}\text{-BPS}} _{SO({\rm even})}(z;t)
&=\sum_ {N=0}^{\infty }Z^{\frac{1}{2}\text{-BPS}}_{SO(2N)}t^N
\nonumber \\
&=\Xi ^{\frac{1}{2}\text{-BPS}}_{SO({\rm odd})}(t;z)
+\Xi ^{\frac{1}{2}\text{-BPS}}_{SO({\rm odd})}(zt;z)
\nonumber\\
&=\Pexp \left (tI^{\ZZ_2}_0(z)\right )
+\Pexp \left (tI^{\ZZ_2}_1(z)\right ),
\label{sopart}
\end{align}
where $I^{\ZZ_2}_m(z)$ ($m=0,1$)
are
defined by
\begin{align}
I^{\mathbb{Z}_2}_0(z)=\frac{1}{1-z^2}
=\frac{I(z)+I(-z)}{2},\quad
I^{\mathbb{Z}_2}_1(z)=\frac{z}{1-z^2}
=\frac{I(z)-I(-z)}{2}.
\label{sopart2}
\end{align}
These are even and odd part of $I(z)\equiv I(0,0,z)$, respectively.
Namely, $I(z)=I^{\mathbb{Z}_2}_0(z)+I^{\mathbb{Z}_2}_1(z)$, and
they satisfy
$I^{\mathbb{Z}_2}_m(-z)=(-1)^mI^{\mathbb{Z}_2}_m(z)$.

%%%%%%%%%%%%%%%%%%%%%%%%%%%%%%%%%%%%%%%%%%%%%%%%%%%%%%%%
\subsection{$\frac{1}{8}$-BPS partition function}\label{sosp8.sec}
We can obtain similar expressions to (\ref{soodd2}) and (\ref{sopart}) for the $\frac{1}{8}$-BPS partition functions
by using the oscillator description.

Let us consider the $SO(2N+1)$ case first.
Similarly to the $U(N)$ case,
we can diagonalize $\mathsf{X}$, $\mathsf{Y}$, and $\mathsf{Z}$ simultaneously by the gauge transformation
so that they become elements of
\begin{align}
so(2)^N\subset so(2N+1),
\label{so2n}
\end{align}
and
eigenvalues $(x_i,y_i,z_i)$ and $(-x_i,-y_i,-z_i)$
associated with the $i$-th $so(2)$ factor
always appear in pair except a single zero
corresponding to the ($2N+1$)-th direction.
Gauge invariant operators,
or functions in the moduli space $\CC^{3N}/{\cal W}_{SO(2N+1)}$,
are expressed as polynomials of $3N$ variables $x_i$, $y_i$, and $z_i$.

We treat a triplet $(x_i,y_i,z_i)$ as the coordinates of a three-dimensional harmonic oscillator.
The Weyl group
${\cal W}_{SO(2N+1)}=S_N\rtimes\ZZ_2^N$
consists of permutations among these $N$ triplets and sign change for an arbitrary $i$,
which is realized by $SO(3)$ rotation acting on the $\RR^3$
consisting of the two directions associated with the $i$-th $so(2)$ factor
in (\ref{so2n}) and the exceptional ($2N+1$)-th direction.
Again, we can regard this as a system of $N$ bosonic particles in the
three-dimensional harmonic potential.
The invariance under the sign change requires the wave function of each particle to be even under the $\mathbb{Z}_2$ action.
Namely, the one-particle wave function $\Psi $ must satisfy
\begin{align}
\Psi (-x,-y,-z) = \Psi (x,y,z).
\label{evenwf}
\end{align}
This condition projects out one-particle states
with odd energy eigenvalues, and the grand partition function is given by
\begin{align}
\Xi_{SO({\rm odd})}(x,y,z;t)=\Pexp \left( tI^{\mathbb{Z}_2}_0(x,y,z) \right),
\label{xiodd}
\end{align}
where
\begin{align}
I^{\ZZ_2}_0(x,y,z)=\frac{1}{2}\left (I(x,y,z)+I(-x,-y,-z)\right ).
\label{iz20}
\end{align}
On the AdS side,
(\ref{iz20}) has a clear interpretation
at least in the large $N$ limit.
We can treat the fugacities $x$, $y$, and $z$ as if they are the
coordinates of $\CC^3$ which contains $\bm{S}^5$.
The $\ZZ_2$ action
\begin{align}
(x,y,z)\rightarrow (-x,-y,-z)
\label{oriz2}
\end{align}
is nothing but the orientifold action, and
(\ref{iz20}) can be regarded as the single-particle partition function
of KK modes
in $\bm{S}^5/\mathbb{Z}_2$.

In the $SO(2N)$ case, we can again use $N$ triplets $(x_i,y_i,z_i)$
to describe gauge invariant operators.
The difference from the $SO(2N+1)$ case is that
the Weyl group of $SO(2N)$ is
${\cal W}_{SO(2N)}=S_N\rtimes\ZZ_2^{N-1}$ and
we cannot change the signs independently for each $i$,
because there is no room to perform the $SO(3)$ rotation
which we used to flip the sign in the $SO(2N+1)$ case.
Although we can change signs of two triplets simultaneously,
it is not possible to change the sign of a single triplet.
This restriction of the sign change
weakens the requirement of the Weyl invariance,
and in addition to (\ref{evenwf})
we have the other solution for the wave function, which satisfies
the twisted boundary condition
\begin{align}
\Psi (-x,-y,-z)=-\Psi (x,y,z).
\label{twistedbc}
\end{align}
The associated projection leaves odd order terms in the Taylor expansion
of the single-particle partition function,
\begin{align}
I^{\mathbb{Z}_2}_1(x,y,z)=\frac{1}{2}\left (I(x,y,z)-I(-x,-y,-z)\right ),
\label{iz21}
\end{align}
and gives the grand partition function $\Pexp \left( tI^{\mathbb{Z}_2}_1 \right)$.
Namely, the $\frac{1}{8}$-BPS grand partition function of the $SO(2N)$ SYM
is the sum of two contributions just like the $\frac{1}{2}$-BPS partition function
(\ref{sopart}),
\begin{align}
\Xi_{SO({\rm even})}(x,y,z;t)
=\Pexp \left( tI^{\mathbb{Z}_2}_0(x,y,z) \right)+\Pexp \left( tI^{\mathbb{Z}_2}_1(x,y,z) \right).
\label{xieven}
\end{align}

The twisted boundary condition
(\ref{twistedbc}) removes
even energy eigenvalues of the harmonic oscillator.
In particular,
oscillators in the
twisted sector cannot be in the zero energy ground state.
All $N$ oscillators contribute at least one unit to the energy,
and the total energy is at least $N$.
Therefore, it is natural to identify
the twisted sector to the contribution of
the Pfaffian type operators including the $\epsilon$ tensor.
This sector decouples in the large $N$ limit,
and is effective only for finite $N$.

How should we interpret this ``twisted sector''
on the gravity side?
It is known that a gauge invariant operator containing the $\epsilon$ tensor
corresponds to a D3-brane wrapped
around the non-trivial three-cycle
in $\bm{S}^5/\mathbb{Z}_2$ \cite{Witten:1998xy}.
The mass of the wrapped D3-brane is $N/(\mbox{AdS radius})$,
and the corresponding operator has dimension $\sim N$.
Therefore, it is natural to identify the twisted sector
to the contribution of a wrapped D3-brane.

We can calculate the partition function
of the $Sp(N)$ theory in a similar way.
The scalar fields $\mathsf{X}$, $\mathsf{Y}$, and $\mathsf{Z}$
are $2N\times 2N$ matrices, and
we can diagonalize them so that they become elements
of $sp(1)^N\subset sp(N)$.
Each $sp(1)$ factor is associated with the coordinates
$(x_i,y_i,z_i)$ of a three-dimensional harmonic oscillator.
Unlike the $SO(2N)$ case we can flip the sign of the coordinates
of each particle independently by the rotation
in the $Sp(1)$ factor, and therefore $\mathcal{W}_{Sp(N)}=\mathcal{W}_{SO(2N+1)}$.
The condition imposed on the wave function
is the same as (\ref{evenwf}) in the $SO(2N+1)$ case,
and the partition function is given by
(\ref{xiodd}).
This is of course the expected result from the Montonen-Olive duality.

As we saw above, the difference between
$\Xi_{SO({\rm even})}$ and
$\Xi_{SO({\rm odd})}=\Xi_{Sp(*)}$
is the choice of sectors summed up.
For unified description of these formulas
we introduce the parameter $p\in\ZZ_2$,
which is $p=0$ for $SO(2N)$ and $p=1$ for $SO(2N+1)$ and $Sp(N)$.
Then, the formulas (\ref{xiodd}) and (\ref{xieven}) are unified into
\begin{align}
\Xi_{S(2,*,p)}(x,y,z;t)=\sum_{pm=0}\Pexp \left(tI^{\ZZ_2}_m(x,y,z) \right).
\label{xit}
\end{align}
The summation is taken over $m\in\{0,1\}$ satisfying
$pm=0$.
Namely, $m=0,1$ for $p=0$ and $m=0$ for $p=1$.

From the viewpoint of the gravity dual,
$p$ is related to the discrete torsion of the three-form fluxes \cite{Witten:1998xy}.
In string theory $SO$ and $Sp$ theories are
realized by using O3-planes.
There are four types of planes
$O3^\pm$ and $\wt{O3}^\pm$, which are
distinguished by an element of the discrete torsion group
associated with the three-form flux fields:
\begin{align}
H^3(\bm{S}^5/\ZZ_2,\wt{\ZZ+\ZZ})=\ZZ_2+\ZZ_2,
\end{align}
where $\wt{\ZZ+\ZZ}$ is the sheaf of
a pair of integers twisted by the orientifold action.
The trivial element corresponds to $p=0$,
and the others to $p=1$.

As a consistency check, we can easily confirm that the
formulas
 (\ref{xiun}) for $U(N)$,
(\ref{zuone}) for $U(1)$,
(\ref{supf}) for $SU(N)$,
(\ref{xiodd}) for $SO(2N+1)$,
and (\ref{xieven}) for $SO(2N)$
are consistent to
Lie algebra isomorphisms.
Namely, the following relations hold.
\begin{align}
Z_{SO(2)}&=Z_{U(1)}, &
Z_{SO(3)}&=Z_{SU(2)}, &
Z_{SO(4)}&=(Z_{SU(2)})^2, &
Z_{SO(6)}&=Z_{SU(4)}.
\end{align}

%%%%%%%%%%%%%%%%%%%%%%%%%%%%%%%%%%%%%%%%%%%%%%%%%%%%%%%%%%%%%
\section{S-fold theories}\label{sfolds.sec}

A $\ZZ_k$ S-fold with $k=3,4,6$ is defined as a
generalization of the orientifold by replacing
the $\ZZ_2$ action (\ref{oriz2})
by the $\ZZ_k$ action which also acts non-trivially on
one of the four supercharges in the ${\cal N}=4$ SYM
\cite{Garcia-Etxebarria:2015wns}.
In the following subsections we study BPS partition functions
of such theories.

Before starting the analysis let us carefully choose the
$\ZZ_k$ action on the scalar fields and comment on the
relations to the Coulomb branch and Higgs branch Hilbert series.
The choice of $\ZZ_k$ is not unique and depends on
the choice of the eliminated supercharge in the reduction from ${\cal N}=4$ to ${\cal N}=3$.
This must be consistent to the definition of the
BPS partition function, in which we need to
use one supercharge to write down the
BPS condition.
Each of the choices of a supercharge
breaks $SU(4)_R$ symmetry to $SU(3)\times U(1)$,
and the scalar fields in the vector representation $\bm{6}$
split into $\bm{3}_{+1}+\ol{\bm{3}}_{-1}$,
``holomorphic'' ones and ``anti-holomorphic'' ones.
Namely, a choice of supercharge fixes a complex structure
in $\RR^6$.
The scalar fields $\sX$, $\sY$, and $\sZ$ are holomorphic
with respect to the complex structure associated with
the supercharge used in the BPS condition.
What is important is that
the supercharge chosen in the construction of the S-fold
should be different from the one used in the BPS condition.
As a consequence the S-fold action on $\sX$, $\sY$, and $\sZ$
cannot be homogeneous.
We adopt the convention with
\begin{align}
(\sX,\sY,\sZ)\rightarrow (\omega_k^{-1}\sX,\omega_k\sY,\omega_k\sZ),\quad
\omega_k &=\exp \left (\frac{2\pi i}{k}\right ).
\end{align}

If we are interested in the Coulomb branch or the Higgs branch Hilbert series
we need to choose an ${\cal N}=2$ subalgebra in the ${\cal N}=3$ algebra.
Again, we need to specify one supercharge from three, that
is not contained in the ${\cal N}=2$ subalgebra.
This again gives a corresponding complex structure,
which is different from the ones appearing above.
This splits three scalar fields $\sX$, $\sY$, and $\sZ$ into
one belonging to an ${\cal N}=2$ vector multiplet and two
belonging to an ${\cal N}=2$ hypermultiplet.
This $1+2$ splitting is different from
that associated with the S-folding.
Namely, the scalar field belonging to the ${\cal N}=2$ vector
multiplet cannot be $\sX$.
If we adopt the convension in which $\sZ$ belongs to the ${\cal N}=2$ vector multiplet
the Coulomb (Higgs) branch Hilbert series is obrained from the
BPS partition function given below by the specialization
$x=y=0$ ($z=0$) as shown in (\ref{specializations}).

%%%%%%%%%%%%%%%%%%%%%%%%%%%%%%%%%%%%%%%%%%%%%%%%%%%%%%%%%%%%%%%%%%%%
\subsection{Grand partition functions for $\mathbb{Z}_{3,4,6}$ S-folds}

Let us generalize the formula (\ref{xit}) to S-folds with $k=3,4,6$.
We use the oscillator representation.
Namely, we express gauge invariant operators as polynomials
which are invariant under ``the Weyl group'' ${\cal W}_{S(k,N,p)}$.
Of course an S-fold theory with $k\geq3$ is not a gauge theory,
and we cannot define the Weyl group
as a subgroup of the gauge group.
However, because an S-fold theory is defined as
the theory on D3-branes, we can define
${\cal W}_{S(k,N,p)}$ as the permutation group of $N$ D3-branes
put in the S-fold background.
Aharony and Tachikawa \cite{Aharony:2016kai}
proposed the action of ${\cal W}_{S(k,N,p)}$ on the Coulomb branch coordinates $z_i$
by generalizing the Weyl groups for $k=1$ and $k=2$ cases.
We simply generalize it to the total moduli space,
and define
${\cal W}_{S(k,N,p)}$ as the group generated by
the following operations.
\begin{align}
&\{(x_i,y_i,z_i),(x_j,y_j,z_j)\}\rightarrow\{(x_j,y_j,z_j),(x_i,y_i,z_i)\},\quad i\neq j,\label{s1}\\
&\{(x_i,y_i,z_i),(x_j,y_j,z_j)\}\nonumber\\
&\quad\rightarrow\{(\omega_k^{-1}x_i,\omega_ky_i,\omega_kz_i),(\omega_k x_j,\omega_k^{-1}y_j,\omega_k^{-1}z_j)\},\quad i\neq j,\label{s2}\\
&(x_i,y_i,z_i)\rightarrow (\omega_k^{-p} x_i,\omega_k^p y_i,\omega_k^p z_i).\label{s3}
\end{align}
The integer $p$ in (\ref{s3}) can be assumed to be a divisor of $k$.
We use $p=0$ instead of $p=k$.
We want to interpret these operations from the viewpoint of
the oscillator description.

We regard a function of $3N$ variables $(x_i,y_i,z_i)$ as the wave function
of an $N$-particle system, and require it to be invariant under
(\ref{s1}), (\ref{s2}), and (\ref{s3}).
The first operation (\ref{s1}) generates permutations of $N$ particles,
and the invariance under this operation can be implemented as
the Bose statistics of the particles,
and the $N$-particle wave function is given in the form (\ref{wavefn}).
The invariance under the second operation requires
all single-particle wave functions satisfy a common
boundary condition
\begin{align}
\Psi (\omega_k^{-1} x,\omega_k y,\omega_k z)&=\omega_k^m\Psi (x,y,z),
\label{wf346}
\end{align}
where the integer $m\in\{0,1,2,\ldots,k-1\}$
is common for all particles, and specifies one of $k$ sectors.
$m=0$ is the untwisted sector and the others are twisted sectors.
The corresponding single-particle partition function is
\begin{align}
I_m^{\mathbb{Z}_k}(x,y,z)
&=\frac{1}{k}\sum_{\ell=0}^{k-1}\frac{\omega_k^{-m\ell}}{(1-\omega_k^{- \ell} x)(1-\omega_k^\ell y)(1-\omega_k^\ell z)}
\nonumber\\
&=\sum_{n_x,n_y,n_z}x^{n_x}y^{n_y}z^{n_z}
,\label{Formula}
\end{align}
where the summation $\sum_{n_x,n_y,n_z}$ is taken over the three integers satisfying
the conditions
\begin{align}
n_x, n_y, n_z\geq0,\quad
-n_x+n_y+n_z=m\mod k.
\label{nxyzcond}
\end{align}
$I_m^{\mathbb{Z}_k}$ satisfies
\begin{align}
I_m^{\mathbb{Z}_k}(\omega_k^{-1}x,\omega _ky,\omega _kz)=\omega _k^mI_m^{\mathbb{Z}_k}(x,y,z),
\label{imrelation}
\end{align}
and reproduces (\ref{iz20}) and (\ref{iz21}) for $k=2$.

In the $m$-th sector with $m\neq0$ the $\ZZ_k$ projection leaves only single-particle states
with non-vanishing energies,
and the lowest energy of the $N$-particle system is of order $N$.
These states are related to the Pfaffian-like operators.
On the gravity side we regard $m$-th sector
as the contribution of
wrapped D3-brane with the winding number $m\in H_3(\bm{S}^5/\ZZ_k,\ZZ)=\ZZ_k$.

The grand partition function is obtained by
summing up
the contribution of sectors.
The contribution of the $m$-th sector
is $\Pexp \left(tI^{\ZZ_k}_m \right)$, and
the sectors summed up are determined
by the invariance under (\ref{s3}).
It requires
$m$ to satisfy the condition
\begin{align}
pm=0\mod k.
\end{align}
Namely, the grand partition function
is given by
\begin{align}
\Xi_{S(k,*,p)}(x,y,z;t)
&=\sum_{pm=0}\Pexp \left(tI^{\ZZ_k}_m(x,y,z) \right)
\nonumber\\
&=\sum_{pm=0}\prod_{n_x,n_y,n_z}\frac{1}{1-tx^{n_x}y^{n_y}z^{n_z}},
\label{xit2}
\end{align}
where the product in the final expression is taken over three integers satisfying
the conditions (\ref{nxyzcond}).
This is the main result in this paper.

As is explained in \cite{Aharony:2016kai},
$p$ is related to the discrete torsion
in the gravity dual,
and it is shown that only $p=0$ or $p=1$ is allowed
for a consistent theory.
The discrete torsion group for the $\ZZ_k$ S-fold is
\begin{align}
\Gamma^{(k)}=H^3(\bm{S}^5/\ZZ_k,\wt{\ZZ+\ZZ}),
\end{align}
where $\wt{\ZZ+\ZZ}$ is the sheaf of a pair of integers twisted by the S-fold action.
For each $k$ this is given by \cite{Aharony:2016kai,Imamura:2016abe}:
\begin{align}
\Gamma^{(2)}&=\ZZ_2+\ZZ_2, &
\Gamma^{(3)}&=\ZZ_3, &
\Gamma^{(4)}&=\ZZ_2, &
\Gamma^{(6)}&=0.
\end{align}
The sectors summed up are
determined by the condition that
the gauge bundle of the corresponding wrapped D3-brane
is consistently defined.
If the discrete torsion is trivial, this is the case for an
arbitrary winding number $m\in H_3(\bm{S}^5/\ZZ_k,\ZZ)=\ZZ_k$,
and this corresponds to $p=0$.
Otherwise, the non-trivial
NS-NS and R-R three-form fluxes
induce electric and magnetic charge on
the worldvolume of the wrapped D3-brane.
This obstructs the definition of the gauge bundle,
and only $m=0$ sector is allowed.
This corresponds to $p=1$.

In the large $N$ limit only the $m=0$ sector contributes to the partition function
and 
(\ref{xit2}) reduces to
\begin{align}
Z_{S(k,\infty,p)}(x,y,z)=\lim_{t\to 1}\left [(1-t)\Xi_{S(k,*,p)} \right ]
=\Pexp \left(I^{\ZZ_k}_0(x,y,z)-1 \right).
\label{sfoldproj}
\end{align}
(The relation (\ref{sfoldproj}) holds not only for the
BPS partition function but also for the superconformal index.
This was used in \cite{Imamura:2016abe} to calculate
the superconformal indices of S-folds in the large $N$ limit.)

%%%%%%%%%%%%%%%%%%%%%%%%%%
\subsection{SUSY enhancement from $\mathcal{N}=3$ to $\mathcal{N}=4$}
It is known that in S-fold theories $S(k,1,0)$ and $S(k,2,0)$ with $k=3,4,6$
the supersymmetry is enhanced from ${\cal N}=3$ to ${\cal N}=4$ \cite{Nishinaka:2016hbw,Aharony:2016kai}.
Let us confirm the consistency of our formula to this phenomenon.
For the trivial discrete torsion $p=0$
the grand partition function is the sum of
the contributions of all sectors $m=0,\ldots,k-1$:
\begin{align}
\Xi_{S(k,*,0)}(x,y,z;t)
&=\sum _{m=0}^{k-1}\Pexp \left(I_m^{\mathbb{Z}_k}(x,y,z)t\right).
\label{eq50}
\end{align}

By picking up $t^1$ terms from (\ref{eq50}) we obtain the partition function
of the rank one S-fold theories
\begin{align}
Z_{S(k,1,0)}(x,y,z)=\sum_{m=0}^{k-1}I^{\ZZ_k}_m(x,y,z)=I(x,y,z),
\end{align}
and this is the partition function of the ${\cal N}=4$ $U(1)$ theory.

The rank two S-fold theories are expected to be equivalent to
${\cal N}=4$ SYM with the gauge groups shown in Table \ref{N3toN4}.
\begin{table}[htb]
\caption{${\cal N}=4$ SYMs expected to be equivalent to $S(k,2,0)$.}\label{N3toN4}
\centering
\begin{tabular}{|c|c|}\hline
S-folds & ${\cal N}=4$ SYMs \\ \hline \hline
$k=3$ &$SU(3)$  \\ \hline
$k=4$ &$SO(5)$  \\ \hline
$k=6$ &$G_2$  \\ \hline
\end{tabular}
\end{table}

The coefficient of $t^2$ term in $\Xi_{S(k,*,0)}$
is
\begin{align}
Z_{S(k,2,0)}(x,y,z)
=\sum_{m=0}^{k-1}\frac{1}{2}
\left[I^{\ZZ_k}_m(x^2,y^2,z^2)
+\left(I^{\ZZ_k}_m(x,y,z)\right)^2\right].
\label{rank2}
\end{align}
For $k=3$ and $k=4$, the
results are
\begin{align}
Z_{S(3,2,0)}&=\frac{1}{6(1-x)^2(1-y)^2(1-z)^2}+\frac{1}{2(1-x^2)(1-y^2)(1-z^2)}\nonumber \\
&\quad +\frac{1}{3(1+x+x^2)(1+y+y^2)(1+z+z^2)},\\
Z_{S(4,2,0)}&=\frac{1}{8(1-x)^2(1-y)^2(1-z)^2}+\frac{1}{8(1+x)^2(1+y)^2(1+z)^2}\nonumber \\
&\quad +\frac{1}{2(1-x^2)(1-y^2)(1-z^2)}+\frac{1}{4(1+x^2)(1+y^2)(1+z^2)}.
\end{align}
These partition functions are the same as those of $\mathcal{N}=4$ $SU(3)$ and $SO(5)$, respectively.
\begin{align}
Z_{S(3,2,0)}(x,y,z)&=Z_{SU(3)}(x,y,z),&
Z_{S(4,2,0)}(x,y,z)&=Z_{SO(5)}(x,y,z).
\label{zs3zs4}
\end{align}

For $k=6$ case 
(\ref{rank2}) gives
\begin{align}
Z_{S(6,2,0)}&=\frac{1}{12(1-x)^2(1-y)^2(1-z)^2}+\frac{1}{12(1+x)^2(1+y)^2(1+z)^2}\nonumber \\
&+\frac{1}{2(1-x^2)(1-y^2)(1-z^2)}+\frac{1}{6(1-x+x^2)(1-y+y^2)(1-z+z^2)}\nonumber \\
&+\frac{1}{6(1+x+x^2)(1+y+y^2)(1+z+z^2)}.
\label{s6}
\end{align}
This is expected to be the same as $Z_{G_2}$.
The specialization of this partition function by $x=y=0$
agrees with the $\frac{1}{2}$-BPS partition function
obtained from the general formula
(\ref{halfbpsz}) with $\{d_1,d_2\}=\{2,6\}$
for $G_2$ shown in Table \ref{Casimir1}.
We can also compare this partition function to
the partition function of the $\ZZ_2$ gauging of the $SU(3)$ SYM.
As is pointed out in \cite{Argyres:2018wxu}
the $\frac{1}{2}$-BPS partition function of the $G_2$ SYM is
obtained from the $SU(3)$ SYM by the discrete gauging of $\ZZ_2$
charge conjugation symmetry.
This is also the case for the $\frac{1}{8}$-BPS partition function,
and the $\frac{1}{8}$-BPS partition function of the $G_2$ theory is given by
\begin{align}
Z_{G_2}(x,y,z)
=\frac{1}{2}
\left[
Z_{SU(3)}(x,y,z)
+Z_{SU(3)}(-x,-y,-z)
\right].
\label{G2}
\end{align}
It is also possible to use (a refinement of)
the Molien series to obtain the same partition function.
We can directly confirm the agreement of
(\ref{s6}) and (\ref{G2}).
\begin{align}
Z_{S(6,2,0)}(x,y,z)&=Z_{G_2}(x,y,z).
\end{align}

(\ref{zs3zs4}) and (\ref{G2})
means that
$Z_{S(3,2,0)}$ and $Z_{S(6,2,0)}$ are related
by the $\ZZ_2$ discrete gauging.
This is explicitly shown in the next subsection.

%%%%%%%%%%%%%%%%%%%%%%%%%%%%%%%%%%%%%%%%%%%%%%%%%%%%%%%%%%%%%
\subsection{Discrete gauging}
The discrete gauging
is the prescription to obtain
a new theory from a parent theory by gauging a discrete symmetry
of the parent theory.
It provides another way to construct ${\cal N}=3$
theories \cite{Argyres:2018wxu,Bourton:2018jwb}.
We remark that the gauging is different from
the S-folding, and S-fold partition functions
are not necessarily obtained from $Z_{U(N)}$
by a gauging.
Instead, the gaugings give additional relations
among S-fold partition functions as we will show
shortly.
Some of the relations below belong to the class
of discrete gaugings associated with the principal
extensions of the gauge groups,
which are investigated in
\cite{Bourget:2018ond}.

As is pointed out in \cite{Aharony:2016kai}
the S-fold theory $S(k,N,p)$ has a $\ZZ_p$ global symmetry.
(In this section we use $p=k$ instead of $p=0$.)
Let $q$ be a divisor of $p$.
We can gauge the subgroup $\ZZ_q\subset\ZZ_p$
to define another theory.
Let us denote the new theory by $S(k,N,p)/\ZZ_q$.
In terms of oscillator variables
$\ZZ_q$ is generated by the rotation
(\ref{s3}) for a single oscillator with $p$ replaced
by $p'=p/q$.
This rotates the total wave function (\ref{wavefn})
by the phase factor $\omega_k^{mp'}$, and the
gauge invariance requires
\begin{align}
p'm=0\mod k.
\label{pmmodk}
\end{align}
As the result, the partition function
of this theory is the same as that
of $S(k,N,p')$:
\begin{align}
Z_{S(k,N,p)/\ZZ_q}
=Z_{S(k,N,p')}.
\label{zzz}
\end{align}
Note that this relation is a reflection of
the coincidence of the moduli space,
and does not mean the equivalence of the two theories.
In general, the two theories are different theories that have
different central charges.
(For example, $S(2,N,2)/\ZZ_2$ is the $O(2N)$ SYM
while $S(2,N,1)$ is the $SO(2N+1)$ SYM.)
The $\frac{1}{2}$-BPS partition function obtained
by the restriction $x=y=0$ is
\begin{align}
Z_{S(k,N,p')}^{\frac{1}{2}\text{-BPS}}
=\Pexp \left (\sum _{j=1}^{N-1}z^{jk}+z^{N(k/p')}\right ),
\end{align}
and is consistent to the
spectrum of Coulomb branch operators given in \cite{Aharony:2016kai}.

We can consider another type of discrete gaugings of S-fold theories
that is generated by $(x_i,y_i,z_i)\rightarrow (\omega_\ell^{-1} x_i,\omega_\ell y_i,\omega_\ell z_i)$
for all oscillators.
$\ell$ may be or may not be a divisor of $k$.
If $\ell$ is a divisor of $k$
this rotation is realized by repeating $N$ times the
rotation of a single oscillator,
and this gauging
is equivalent to the
previous gauging with
\begin{align}
p'=N(k/\ell)\mod k.
\label{ppandell}
\end{align}

We can describe the overall rotation
of the all oscillators
by the rotation of the fugacities
\begin{align}
(x,y,z)\rightarrow(\omega_\ell^{-1} x,\omega_\ell y,\omega_\ell z),
\label{zellaction}
\end{align}
and the partition function
of the gauged theory is
\begin{align}
{\cal P}_{\ZZ_\ell}Z,
\label{zdg}
\end{align}
where $Z$ is the partition function
before the $\ZZ_\ell$ projection
and
${\cal P}_{\ZZ_\ell}$ is the projection operator that eliminates
terms that are not invariant under (\ref{zellaction}).
This is in contrast to the S-fold projection (\ref{sfoldproj})
in the large $N$ limit in which the projection is carried out 
before the plethystic exponential.

Let us first confirm if $\ell$ is a divisor of $k$
(\ref{zdg}) reproduces (\ref{zzz})
with $p'$ given by (\ref{ppandell}).
Let ${\cal R}_\ell$ be the operator replacing $(x,y,z)$
by $(\omega_\ell^{-1} x,\omega_\ell y,\omega_\ell z)$.
If $\ell$ is a divisor of $k$
the relation (\ref{imrelation}) holds.
Namely,
${\cal R}_\ell I^{\ZZ_k}_m=\omega_\ell^mI^{\ZZ_k}_m$.
This relation means that
$tI^{\ZZ_k}_m(x,y,z)$ is invariant under
the $\ZZ_\ell$ action
$(x,y,z;t)\rightarrow(\omega_\ell^{-1} x,\omega_\ell y,\omega_\ell z;\omega_\ell^{-m}t)$,
and so is the plethystic exponential
$\Pexp \left(tI^{\ZZ_k}_m\right)$.
Therefore, the coefficient of the $t^N$ term
in the Taylor expansion of $\Pexp\left(tI^{\ZZ_k}_m\right)$ satisfies
\begin{align}
{\cal R}_\ell\left (\Pexp \left.\left(tI^{\ZZ_k}_m\right) \right|_{t^N}\right )
=\omega_\ell^{mN}\Pexp\left.\left(tI^{\ZZ_k}_m\right)\right|_{t^N}.
\label{rkaction}
\end{align}
The ${\cal R}_\ell$ invariance requires $mN=0\mod\ell$,
and this is the same as
the condition (\ref{pmmodk}) with $p'$ given by (\ref{ppandell}).

By using (\ref{rkaction}) we can prove the relation
${\cal P}_{\ZZ_2}Z_{S(3,2,0)}=Z_{S(6,2,0)}$, 
which we mentioned at the end of the previous subsection, as follows.
First we divide the single particle partition function
$I^{\ZZ_3}_m$ into
two parts; $I^{\ZZ_3}_m=I^{\ZZ_6}_m+I^{\ZZ_6}_{m+3}$.
With this decomposition we can rewrite the
grand partition function as
\begin{align}
\Xi_{S(3,*,0)}
&=
\sum_{m=0}^2
\Pexp \left(tI^{\ZZ_6}_m \right)
\Pexp \left(tI^{\ZZ_6}_{m+3} \right),
\end{align}
and by picking up $t^2$ terms we obtain
\begin{align}
Z_{S(3,2,0)}
&=
\sum_{m=0}^2\sum_{r=0}^2
\Pexp \left.\left(tI^{\ZZ_6}_m \right)\right|_{t^r}
\Pexp \left.\left(tI^{\ZZ_6}_{m+3} \right)\right|_{t^{2-r}}.
\end{align}
When we apply ${\cal R}_2$
the summand is rotated by the
phase
$\omega_2^{rm}\omega_2^{(2-r)(m+3)}=(-1)^r$,
and the corresponding projection
${\cal P}_{\ZZ_2}$
leaves the terms with $r=0$ and $r=2$.
We obtain
\begin{align}
{\cal P}_{\ZZ_2}Z_{S(3,2,0)}
&=
\sum_{m=0}^2\left[
\Pexp \left.\left(tI^{\ZZ_6}_{m+3} \right)\right|_{t^2}
+
\Pexp \left.\left(tI^{\ZZ_6}_m \right)\right|_{t^2}\right]
=Z_{S(6,2,0)}.
\end{align}
This is the relation we wanted to show.

%%%%%%%%%%%%%%%%%%%%%%%%%%%%%%%%%%%%%%%%%%%%%%%%%%%%%%%%%%
%%%%%%%%%%%%%%%%%%%%%%%%%%%%%%%%%%%%%%%%%%%%%%%
\section{D3-brane analysis}\label{d3.sec}
In this section we reproduce the partition function (\ref{xit2})
by quantizing D3-branes in $\bm{S}^5/\ZZ_k$.
The analysis is quite similar to the analysis of sphere giants in \cite{Biswas:2006tj}.
Actually we can use the essential part of the calculation in \cite{Biswas:2006tj} as it is for our purpose.
The analysis in \cite{Biswas:2006tj} starts from the BPS brane configuration obtained
by Mikhailov \cite{Mikhailov:2000ya}.
Mikhailov showed that an arbitrary BPS solution is given as the intersection of
$\bm{S}^5$ defined by $|X|^2+|Y|^2+|Z|^2=1$ and a holomorphic surface $f(X,Y,Z)=0$.
We consider the Taylor expansion
\begin{align}
f(X,Y,Z)=\sum_{n_x,n_y,n_z}c_{n_x,n_y,n_z}X^{n_x}Y^{n_y}Z^{n_z},
\label{expans}
\end{align}
and treat the coefficients $c_{n_x,n_y,n_z}$ as dynamical variables.
Because the overall factor of $f$ is irrelevant to the brane configuration
the coefficients are regarded as the projective coordinates of $\CC\bm{P}^\infty$.
Due to the coupling of the D3-brane to the background RR flux
the wave function is not just a function but a section of the line bundle ${\cal O}(N)$ over this
configuration space.
Therefore, the quantization reduces to the simple problem to find holomorphic sections of this line bundle.
There are two issues which make the problem complicated.
One is that different functions $f$ may give the same brane configuration, and we should remove the redundancy.
The other is that the surface $f=0$ may not intersect with $\bm{S}^5$,
and the parameter region giving such a surface should be removed from the configuration space.
The detailed analysis in \cite{Biswas:2006tj} shows that
even if we take account of these issues the result is the same as what we obtained by naive analysis
neglecting these issues.

Let us assume that this is the case for the S-fold. 
Then what we should additionally do is to impose the $\ZZ_k$ invariance to the surface $f=0$.
This requires the function satisfy
\begin{align}
f(\omega_k^{-1}X,\omega_kY,\omega_kZ)=\omega_k^m f(X,Y,Z),
\label{zkf}
\end{align}
with some $m\in\ZZ_k$.
We identify $m$ with the winding number
of a D3-brane around the non-trivial cycle in $\bm{S}^5/\ZZ_k$.
This is easily shown as follows.
We can deform the function $f$ by continuously changing coefficients
to a simple function, say, $f=Z^m$, without violating the property (\ref{zkf}).
The resulting configutation obviously has the winding number $m$.
Because such a deformation does not change the homology class of the brane configuraion,
an arbitrary brane configuration given by a function satisfying (\ref{zkf}) has winding number $m$.

Now, let us follow the quantization procedure of \cite{Biswas:2006tj} under the restriction (\ref{zkf}).
The configuration space is again $\CC\bm{P}^\infty$ with the homogeneous coordinates
$c_{n_x,n_y,n_z}$.
The constraint (\ref{zkf}) requires $(n_x,n_y,n_z)$ to
satisfy
\begin{align}
-n_x+n_y+n_z=m \mod k,
\label{pqrm}
\end{align}
and this is the same as (\ref{nxyzcond}).
The wave function is a holomorphic section of the ${\cal O}(N)$ line bundle over this
configuration space, and is given by an order $N$ homogeneous function of $c_{n_x,n_y,n_z}$.
We can treat each of $c_{n_x,n_y,n_z}$ as if it is a quantum with angular momentum $(J_1,J_2,J_3)=(n_x,n_y,n_z)$.
Then a quantum state of D3-branes in $\bm{S}^5$ is regarded as a collection of $N$ quanta,
and the partition function of D3-branes can be calculated as the partition function of
states which include $N$ quanta.
Let us introduce $(x,y,z)$ and $t$ as fugacities for the angular momenta $(J_1,J_2,J_3)$ and
the number of quanta $N$, respectively, and calculate the grand partition function.
The contribution of a single quantum of $c_{n_x,n_y,n_z}$ is $tx^{n_x}y^{n_y}z^{n_z}$,
and the grand partition function for a fixed winding number $m$ is
\begin{align}
\Pexp\left(\sum_{n_x,n_y,n_z}tx^{n_x}y^{n_y}z^{n_z}\right),
\label{zm}
\end{align}
where the sum is taken over non-negative integers $(n_x,n_y,n_z)$ satisfying
(\ref{pqrm}).
By summing up (\ref{zm}) over $m$ allowed by the discrete torsion,
we obtain the grand partition function $\Xi_{S(k,*,p)}(x,y,z;t)$ in (\ref{xit2}).

%%%%%%%%%%%%%%%%%%%%%%%%%%%%%%%%%%%%%%%%%%%%%%%%%%%%%%%%%%
\section{Discussion}\label{discussion.sec}
In this paper we derived the BPS partition functions
for arbitrary S-fold theories.
We confirmed that the formula is consistent to the Lie algebra isomorphisms.
It is also consistent to the supersymmetry
enhancement from $\mathcal{N}=3$ to $\mathcal{N}=4$ in rank $1$ and $2$ theories.
Namely, for $S(k,1,0)$ the partition function is the same as the $U(1)$ SYM,
and for $S(k,2,0)$ with $k=3,4,6$ the partition functions
are the same as those of SYM with $G=SU(3)$, $SO(5)$, and $G_2$, respectively.
We also gave some relations among partition functions via discrete gaugings.

The formula gives the partition function as the sum of contributions
of sectors.
From the holographic point of view,
different sectors correspond to different winding numbers of D3-branes
around the non-trivial cycle in the internal space $\bm{S}^5/\ZZ_k$.
We derived the same formula by quantizing D3-branes in $\bm{S}^5/\ZZ_k$
following the similar analysis of sphere giants \cite{Biswas:2006tj}.

The derivation on the SCFT side is based on the harmonic oscillator description of BPS operators.
In the large $N$ limit the sectors with wrapped branes decouple, and only the untwisted sector contributes 
to the partition function.
Each excited state of a harmonic oscillator
can be regarded as a KK mode in $\bm{S}^5/\ZZ_k$.
For finite $N$, this correspondence is not so obvious.
In particular, the twisted sector gives a Pfaffian operator
as ``a bound state'' of $N$ harmonic oscillators.
Naively, this may be interpreted on the gravity side
that a D3-brane wrapped on the non-trivial cycle in $\bm{S}^5/\ZZ_k$
is a bound state of  KK modes satisfying
the twisted boundary condition.
However, it is not possible to impose the twisted boundary
condition on KK modes due to the absence of gauge fields
minimally coupling to gravitons.
At present, unfortunately, we have no clear explanation
how this is realized.
It may be interesting to study the relation between
the harmonic oscillator description
and the quantization procedure of D3-branes
we used in section \ref{d3.sec}.

Other than the BPS partition function,
there is another important quantity reflecting the operator spectrum:
the superconformal index.
It has many connections to physical quantities.
In particular, its Schur limit (the Schur index)
are known to be related to the BPS spectrum on the Coulomb branch
\cite{Cordova:2015nma},
2d chiral algebra \cite{Beem:2013sza},
and correlation functions in 2d topological QFT
\cite{Gadde:2009kb,Gadde:2011ik}.
Furthermore, there is an analytic formula for the Schur index for $U(N)$ SYM
with an arbitrary $N$ \cite{Bourdier:2015wda}.
It would be very interesting to investigate
the relation of our analysis to the superconformal index.

\section*{Acknowledgments}
We are grateful to T. Mori for wonderful discussions. The work of S.~F. was partially supported by
Advanced Research Center for Quantum Physics and
Nanoscience, Tokyo Institute of Technology.
The work of Y.~I. was 
 partially supported by Grand-in-Aid for Scientific Research (C) (No.15K05044),
Ministry of Education, Science and Culture, Japan.

\appendix
\section{The Molien series}\label{MS}
As we mentioned in the main text
the BPS partition function of an ${\cal N}=4$ SYM
can be calculated by the Molien series.
In this appendix we
show that
the grand partition function (\ref{xiun}) for the
unitary gauge groups is also
obtained by using the Molien series.

Let $V$ be a $d$-dimensional vector space
and $G$ be an orbifold group
acting on $V$.
Let $v_i$ ($i=1,\ldots,d$) be affine coordinates
of $V$,
and $D_{ij}$ be the corresponding $d\times d$ matrix representation
of $G$.
We consider the orbifold defined by the orbifold action
\begin{align}
v_i\rightarrow v_i'=D_{ij}(g)v_j,\quad g\in G.
\label{gv}
\end{align}
The coordinate ring of the orbifold $V/G$ is spanned by
polynomials of $v_i$ invariant under (\ref{gv}).
Let $c_n$ be the number of linearly independent
$G$-invariant homogeneous polynomials of degree $n$.
The generating function $M(z)=\sum_{n=0}^\infty c_nz^n$ is
called ``the Molien series,'' and given by the
formula
\begin{align}
M(z)
=\frac{1}{|G|}\sum _{g\in G}\frac{1}{\det (\mathbb{I}_d-zD(g))}.
\label{molien}
\end{align}
This formula can be applied to an arbitrary orbifold.
In the following we confirm that
the grand partition function
$\Xi_{U(*)}$
can also be derived by using this formula.

Let us first consider $\frac{1}{2}$-BPS partition function.
The Coulomb branch moduli space $\CC^N/S_N$, and
hence the $\frac{1}{2}$-BPS partition function of $U(N)$ SYM
should be obtained by applying the formula
(\ref{molien}) to this orbifold.
Namely, by taking the vector space $V=\CC^N$ and the orbifold group $G=S_N$
(\ref{molien}) gives the $\frac{1}{2}$-BPS partition function
$Z^{\frac{1}{2}\BPS}_{U(N)}(z)$.

Because the summand takes the same value
for permutations in the same class
we can replace
the summation over permutations $g\in S_N$ by
the summation over conjugacy classes with
appropriate multiplicities inserted.
We have
\begin{align}
Z_{U(N)}^{\rm Molien}(z)=\sum _{\mu\in C_N}\frac{m(\mu)}{N!}
\frac{1}{\det (\mathbb{I}_N-zD(g(\mu)))},
\label{ccsum}
\end{align}
where $C_N$ is the family of conjugacy classes of $S_N$,
$m(\mu)$ is the number of elements in a
conjugacy class $\mu\in C_N$,
and $g(\mu)$ is a representative of $\mu$.
A conjugacy class $\mu$ is uniquely specified
by the list of the lengths of cycles in $g(\mu)$.
Let $k_j$ ($j=1,2,\ldots$) be the number of $j$-cycles
in $g(\mu)$.
These satisfy
\begin{align}
\sum_{j=1}^\infty jk_j=N,
\label{totaln}
\end{align}
and the weight factor in
(\ref{ccsum}) is given by
\begin{align}
\frac{m(\mu)}{N!}=
\prod_{j=1}^\infty\frac{1}{j^{k_j}k_j!}.
\end{align}
Note that for a specific $N$
only finite number of $k_j$ are
non-vanishing due to the constraint (\ref{totaln}).
Corresponding to the cycle decomposition,
the matrix $\mathbb{I}_N-zD(g(\mu ))$ appearing in
(\ref{ccsum}) takes the block-diagonal form.
Each block corresponds to each cycle.
The block associated with a $j$-cycle is
\begin{align}
I_j=\begin{pmatrix}1&-z&&&\\ &1&-z&&\\ &&\ddots &\ddots &\\ &&&1&-z\\ -z&&&&1 \end{pmatrix},
\end{align}
and $\det I_j=1-z^j$.
Therefore, we can rewrite (\ref{ccsum}) as
\begin{align}
Z_{U(N)}^{\rm Molien}(z)=\sum _{\{k_j\}}\prod_{j=1}^\infty
\frac{1}{j^{k_j}k_j!}\frac{1}{(1-z^j)^{k_j}},
\label{zmz}
\end{align}
where the sum is taken over $\{k_j\}$ satisfying
(\ref{totaln}).

It is straightforward to extend the analysis above
to the $\frac{1}{8}$-BPS partition function.
We define three copies of $N$-dimensional vector
spaces $V_x$, $V_y$, $V_z$ corresponding to the three
scalar fields
and replace the vector space $V$ by the direct product
$V_x\times V_y\times V_z$,
and correspondingly replace $D(g)$ by $\mathbb{I}_3\otimes D(g)$.
As the result, we obtain
the Molien series for the
full moduli space $\CC^{3N}/S_N$,
which is given by (\ref{zmz}) with the factor
$1/(1-z^j)^{k_j}$ replaced by
$1/(1-z^j)^{3k_j}$.
This is the $\frac{1}{8}$-BPS partition function with
fugacities $x=y=z$.
To obtain the partition function with generic fugacities
we need to consider a refinement of the Molien series \cite{Benvenuti:2006qr}.
It is defined by using
$\diag(x,y,z)\otimes D(g)$ instead of $z\mathbb{I}_3\otimes D(g)$.
Namely,
\begin{align}
Z_{U(N)}^{\rm Molien}(x,y,z)
=\frac{1}{N!}\sum _{g\in S_N}\frac{1}{\det (\mathbb{I}_{3N}-\diag(x,y,z)\otimes D(g))}.
\end{align}
By repeating the same procedure as above we obtain
\begin{align}
Z_{U(N)}^{\rm Molien}(x,y,z)
=\sum _{\{k_j\}}\prod_{j=1}^\infty
\frac{1}{j^{k_j}k_j!}I(x^j,y^j,z^j)^{k_j},
\label{zmxyz}
\end{align}
where $I(x,y,z)$ is the function defined in
(\ref{iustar}).
To obtain the grand partition function,
we multiply $t^N=\prod_{j=1}^\infty (t^j)^{k_j}$
to (\ref{zmxyz}).
\begin{align}
t^NZ_{U(N)}^{\rm Molien}(x,y,z)
=\sum _{\{k_j\}}\prod_{j=1}^\infty
\frac{1}{k_j!}\left(\frac{1}{j}I(x^j,y^j,z^j)t^j\right)^{k_j}.
\end{align}
Finally,
the summation over
$N=0,1,\ldots$
gives the grand partition function.
This summation is equivalent to
removing the constraint (\ref{totaln})
in the summation with respect to $\{k_i\}$.
Then all components in the series $\{k_i\}$ become
independent, and the result becomes
\begin{align}
\prod_{j=1}^{\infty }\sum _{k=0}^{\infty }\frac{1}{k!}
\left (\frac{1}{j}I(x^j,y^j,z^j)t^j\right )^k.
\end{align}
This is the same as the grand partition function
$\Xi_{U(*)}$ given in
(\ref{xiun}).

\end{document}